\documentclass{article}
\usepackage{mathtools}
\usepackage{graphicx}
\usepackage{amsmath}
\usepackage{amssymb}
\usepackage{amsthm}
\usepackage{iftex}
\usepackage{lmodern}
\usepackage{textcomp}
\usepackage{multirow}
\usepackage[symbol]{footmisc}

\theoremstyle{thmstyleone}
\newtheorem{theorem}{Theorem}[section]
\newtheorem{proposition}{Proposition}[section]
\newtheorem{example}{Example}[section]%
\newtheorem{remark}{Remark}[section]%
\newtheorem{lemma}{Lemma}[section]%
\newtheorem{definition}{Definition}[section]%
\newtheorem{corollary}{Corollary}[section]%

\begin{document}
	\begin{center}
		\textbf{\huge Reversible cyclic codes over finite chain rings}
	\end{center}
	\begin{center}
		Monika Dalal$^{1}$, Sucheta Dutt$^{1*}$ and Ranjeet Sehmi$^{1}$\\
		$^{1}$Department of Applied Sciences,Punjab Engineering College\\
		(Deemed to be University), Chandigarh, India, 160012\\
		$^{*}$Corresponding author(s). E-mail(s): $rsehmi@pec.edu.in$ ;\\
		Contributing authors : $monika.phdappsc@pec.edu.in$;\\
		$sucheta@pec.edu.in$; 
	\end{center}

\section*{Abstract}\label{sec0}

       In this paper, necessary and sufficient conditions for the reversibility of a cyclic code of arbitrary length over a finite commutative chain ring have been derived. MDS reversible cyclic codes having length $p^{s}$ over a finite chain ring with nilpotency index 2 have been characterized and a few examples of MDS reversible cyclic codes have been presented. Further, it is shown that the torsion codes of a reversible cyclic code over a finite chain ring are reversible. Also, an example of a non-reversible cyclic code for which all its torsion codes are reversible has been presented to show that the converse of this statement is not true. The cardinality and Hamming distance of a cyclic code over a finite commutative chain ring have also been determined.\\\

       \textbf{Keywords :} Cyclic codes, Reversible codes, Torsion codes, Generators\\

\section{Introduction}\label{sec1}

        The class of cyclic codes is one of the most significant class of codes when it comes to easy encoding and decoding of data. Prange had introduced cyclic codes over finite fields in \cite{bib6}. Cyclic codes over rings have been in focus since Calderbank et al. \cite{bib3} established the fact that some good non linear codes having excellent error correcting capabilities can be constructed as binary images of linear codes over $Z_{4}$ via Gray maps. T. Abualrub and R. Oehmke have studied cyclic codes over $Z_{4}$ and given their structure in \cite{bib15}. A. Garg and S. Dutt \cite{bib2} have given structure of cyclic codes having length $p^{s}$ over $Z_{p^{m}}$. J. Kaur et al. \cite{bib8} have obtained a set of generators for cyclic codes over Galois rings, using the minimal degree polynomials of certain subsets of the code. These results have been extended for cyclic codes having arbitrary length over the class of finite chain rings by Monika et al. \cite{bib11}. A. Sharma and T. Sidana \cite{bib4} have given a unique generating set for $\lambda$-constacyclic codes having length $p^{s}$ over a finite chain ring having nilpotency index $\nu$.

        Reversible codes is another useful class of codes having a significant role in DNA computing, retrieval systems and data storage. For a fixed length, these codes have the capability to detect and correct more number of errors as compared to cyclic codes of the same length. In the case of reversible codes, the process of decoding is much faster as the decoder which is used to decode a codeword $c$ can be used to decode the reverse codeword of $c$ as well, eventually resulting in higher rate of transmission.  Reversible codes over fields were first studied by J. L. Massey \cite{bib9}. T. Abualrub and I. Siap \cite{bib14} have given a condition which is necessary as well as sufficient for a cyclic code over $Z_{4}$ to be reversible. H. Islam and Om Prakash \cite{bib10} and J. Kaur et al. \cite{bib7} have respectively studied reversibility of cyclic codes over $Z_{p^{m}}$ and over Galois rings.

        The focus of the present paper is on reversible cyclic codes and torsion codes of cyclic codes over finite chain rings. The manuscript is organised as follows : In section 2, basic definitions and preliminary results on finite chain rings, cyclic codes, torsion codes of cyclic codes and reversible codes have been recalled. In section 3, a condition which is necessary as well as sufficient for a cyclic code having length $p^{s}$ over a finite chain ring of nilpotency index 2, to be a reversible cyclic code has been established. We have also characterized MDS reversible cyclic codes having length $p^{s}$ over a finite chain ring of nilpotency index 2. A few examples of MDS reversible cyclic codes have been presented. In section 4, necessary and sufficient conditions for the reversibility of a cyclic code of arbitrary length over a finite chain ring have been derived. Further, it is shown that the torsion codes of a reversible cyclic code over a finite chain ring are reversible. Also, an example of a non-reversible cyclic code for which all its torsion codes are reversible has been presented to show that the converse of this statement is not true. The cardinality and Hamming distance of a cyclic code of arbitrary length over a finite chain ring have also been determined.

\section{Preliminaries}\label{sec2}

        Let $R$ be a finite commutative chain ring. Let $\langle \gamma \rangle$  be the unique maximal ideal of $R$ and $\nu$ be the nilpotency index of $\gamma$. Let $F_{q}=R/\langle \gamma \rangle$ be the residue field of $R$, where $q=p^{m}$ for a prime $p$ and a positive integer $m$.\\

        The following is a well known result. For reference, see \cite{bib1} and \cite{bib5}.

       \begin{proposition}
        Let $R$ be a finite commutative chain ring. Then\\
        $(i)$ $charR=p^{a}$, where $1 \leq a \leq \nu$ and $\lvert R \rvert=\lvert F_{q}\rvert^{^{\nu}}=p^{m\nu}$.\\
        $(ii)$ There exists an element $\zeta \in R$ with multiplicative order $p^{m}-2$. The set $\top=\{0,1,\zeta,\zeta^{^{2}},\cdots,\zeta^{^{p^{m}-1}}\}$ is called the Teichm$\ddot{u}$ller set of $R$.\\
        $(iii)$ Every $r \in R$ can be uniquely expressed as $r=r_{_{0}} + r_{_{1}}\gamma + \cdots + r_{_{\nu-1}}\gamma^{_{\nu-1}}$, where $r_{_{i}} \in \top$ for $0 \leq i \leq \nu-1$. Also, $r$ is a unit in $R$ if and only if $r_{_{0}} \neq 0$.
       \end{proposition}

        Define a map $\textendash : R \rightarrow R/\langle \gamma \rangle$ by $r \mapsto r(mod \gamma)$ for $r \in R$. Clearly $\textendash$ is a natural onto homomorphism and therefore $\overline{R}=R/\langle \gamma \rangle$, where $\overline{R}$ denotes the image of $R$ under $\textendash$. This map can be naturally extended from $R[z]$ to $\overline{R}[z]$ by $\Sigma_{i=0}^{k}a_{_{i}}z^{i} \mapsto \Sigma_{i=0}^{k}\overline{a_{_{i}}}z^{i}$, where $a_{i} \in R$ for $0 \leq i \leq k$. Using Proposition 2.1(iii) and the fact that the map $\textendash$ restricted to $\top$ is a bijection, it can be observed that each element of $R$ corresponds to a unique element of $\overline{R} + \gamma \overline{R} + \cdots + \gamma^{_{\nu-1}} \overline{R}$, i.e., $R \cong F_{q} + \gamma F_{q} + \cdots + \gamma^{_{\nu-1}} F_{q}$.\\

       Let us now recall some basic definitions and known results.
       
       \begin{definition}
        A linear code $C$ with length $n$ over a finite commutative chain ring $R$ is said to be a cyclic code if $(c_{_{n-1}},c_{_{0}},\cdots,c_{_{n-2}}) \in C$ for every $(c_{_{0}},c_{_{1}},\cdots,c_{_{n-1}}) \in C$. It is well established that $C$ can be viewed as an ideal of $R[z]/\langle z^{n}-1 \rangle$.
       \end{definition}

       \begin{definition}
     	The Hamming weight $w_{_{H}}(c)$ of $c=(c_{_{0}}, c_{_{1}} \cdots c_{_{n-1}}) \in C$ is defined as the number of integers $i$ such that $c_{_{i}}\neq 0$ for $0 \leq i \leq n-1.$
       \end{definition}

       \begin{definition}
     	The Hamming distance $d_{_{H}}(C)$ of a code $C$ over $R$ is given by $d_{_{H}}(C)=min\{w_{_{H}}(c) : c ~is ~a ~non$-$trivial ~element ~of ~C \}.$
       \end{definition}
     
       \begin{definition}
     	A code $C$ over a ring $R$ is said to be an MDS code with respect to the Hamming metric if $\lvert C\lvert=\lvert R\lvert^{n-d_{H}(C)+1}.$
       \end{definition}

       \begin{definition}
        Let $C$ be a cyclic code having length $n$ over $R$. The $i$-th torsion code of $C$ is defined as Tor$_{_{i}}(C)=\{\overline{k(z)} \in \overline{R}[z] : \gamma^{i}k(z) \in C\},$ where $0 \leq i \leq \nu-1.$
       \end{definition}

       \begin{lemma}
       	Let $C$ be a cyclic code over $R.$ Then Tor$_{_{i}}(C)$ is a principally generated cyclic code over the residue field $F_{p^{m}},$ for all $i,$ $0 \leq i \leq \nu-1.$
       \end{lemma}

      \begin{definition}
      	 The degree of the generator polynomial of Tor$_{_{i}}(C)$ for a cyclic code $C$ over $R$ is called the $i$-th torsional degree of $C$.
      \end{definition}

       \begin{definition}
        A cyclic code $C$ having length $n$ over a ring $R$ is said to be reversible if $(\mathtt{c}_{_{n-1}},\mathtt{c}_{_{n-2}},\cdots,\mathtt{c}_{_{0}}) \in C$ for every $(\mathtt{c}_{_{0}},\mathtt{c}_{_{1}},\cdots,\mathtt{c}_{_{n-1}}) \in C$.
       \end{definition}

       \begin{definition}
        The reciprocal polynomial of a polynomial $k(z) \in R[z]$ is defined by $k^{*}(z)=z^{deg\big(k(z)\big)} k(1/z)$.
       \end{definition}

       \begin{lemma}[Remark 3.2.2,\cite{bib7}]
        For any polynomial $k(z) \in R[z]$, $deg\big(k^{*}(z)\big) \leq deg\big(k(z)\big)$, and equality holds if the constant term of $k(z)$ is non zero.
       \end{lemma}

      \begin{definition}
       A polynomial $k(z) \in R[z]$ is said to be self reciprocal if and only if $k^{*}(z)=k(z)$.
      \end{definition}

      \begin{lemma}[Lemma 19,\cite{bib16}]
       Let $f(z)$ and $g(z)$ be polynomials in $R[z]$ with $deg\big(f(z)\big) \geq deg\big(g(z)\big)$. Then\\
       $(i)$ $\big(f(z) g(z)\big)^{*}=f^{*}(z) g^{*}(z)$,\\
       $(ii)$ $\big(f(z) + g(z)\big)^{*}=f^{*}(z) + z^{deg\big(f(z)\big)-deg\big(g(z)\big)} g^{*}(z)$.
      \end{lemma}

      \begin{lemma}[Theorem 3.3.3,\cite{bib7}]
       Let $C=\langle f_{_{1}}(z),f_{_{2}}(z),\cdots,f_{_{k}}(z) \rangle$ be a cyclic code over $R.$ Then $C$ is reversible if and only if $f^{*}_{_{i}}(z) \in C$ for $1 \leq i \leq k$.
      \end{lemma}

\section{Reversible cyclic codes having length $p^{s}$ over $R$}\label{sec3}
   
        In this section, we establish a condition which is necessary as well as sufficient for the reversibility of a cyclic code having length $p^{s}$ over a finite commutative chain ring of nilpotency index 2. We have characterized  MDS reversible cyclic codes having length $p^{s}$ over a finite chain ring of nilpotency index 2. Lastly, a few examples of cyclic codes have been presented in support of our results.\\

        Throughout this section, let $R$ be a finite commutative chain ring having nilpotency index 2, i.e., $R \cong F_{p^{m}} + \gamma F_{p^{m}}$. Let $C$ be a cyclic code having length $p^{s}$ over $R$. A. Sharma and T. Sidana have given a unique generating set for a $\lambda$-constacyclic code having length $p^{s}$ over a finite chain ring of nilpotency index $\nu$ in Theorem III.4 of \cite{bib4}. Substituting $\lambda=1$ and $\nu=2$ in Theorem III.4 of \cite{bib4}, we obtain a unique generating set for a cyclic code having length $p^{s}$ over $R$ as stated below.

      \begin{lemma}
        Let $R \cong F_{p^{m}} + \gamma F_{p^{m}}$ be a finite commutative chain ring with $\nu=2$. Then all distinct non-zero cyclic codes having length $p^{s}$ over $R$ are\\
        $(i)$ $\langle \gamma (z-1)^{b} \rangle$ with $T_{_{0}}=p^{s}$ and $T_{_{1}}=b$, where $T_{_{0}}$ and $T_{_{1}}$ are the zeroth and first torsional degrees of $C$.\\
        $(ii)$ $\langle (z-1)^{a} + \gamma (z-1)^{t} g(z), \gamma (z-1)^{b} \rangle$ with $T_{_{0}}=a$ and $T_{_{1}}=b$,\\
        where $0 \leq b \leq a \leq p^{s}-1$, $g(z) \in F_{p^{m}}[z]$ is either zero or unit in $R[z]/\langle z^{p^{s}}-1\rangle$ with $deg\big(g(z)\big) < b-t$ and $0 \leq t < b$ if $g(z) \neq 0$.
      \end{lemma}

        In the following theorems 3.1 and 3.2, we establish a necessary as well as sufficient condition for reversibility of a cyclic code having length $p^{s}$ over $R$ using its structure as given in Lemma 3.1.

      \begin{theorem}
   	    Let $C=\langle \gamma (z-1)^{b} \rangle$, $0 \leq b \leq p^{s}-1$ be a cyclic code having length $p^{s}$ over $R$. Then $C$ is a reversible cyclic code.
      \end{theorem}

      \begin{proof}
   	    Consider the reciprocal polynomial of the generator of $C$, i.e., $$\big(\gamma (z-1)^{b}\big)^{*} = \gamma z^{b} \bigg(\frac{1}{z}-1\bigg)^{b} = (-1)^{b} \gamma (z-1)^{b},$$ which clearly belongs to $C$. Therefore, by Lemma 2.4, $C$ is reversible.
   	  \end{proof}

      \begin{theorem}
   	    Let $C=\langle (z-1)^{a} + \gamma (z-1)^{t} g(z), \gamma (z-1)^{b} \rangle$ be a cyclic code having length $p^{s}$ over $R$, where $0 \leq b \leq a \leq p^{s}-1$, $g(z) \in F_{p^{m}}[z]$ is either zero or unit in $R[z]/\langle z^{p^{s}}-1\rangle$ with $deg\big(g(z)\big) < b-t$ and  $0 \leq t < b$ if $g(z) \neq 0$. Then $C$ is reversible if and only if $$(z-1)^{b-t} \lvert \big(z^{a-t-k} B(z) - A(z)\big),$$ where $k=deg\big(g(z)\big)$, $A(z)=(-1)^{a} g(z)$ and $B(z)=(-1)^{t} g^{*}(z)$.
   	  \end{theorem}

      \begin{proof}
   	    It is easy to see that $C$ is reversible when $g(z)=0$.\\
        Now let $g(z) \not =0$. Let $C$ be a reversible cyclic code over $R$. It follows from Lemma 2.4 that $\big((z-1)^{a} + \gamma (z-1)^{t} g(z)\big)^{*} \in C$. So, there exist polynomials $l_{_{1}}(z)$ and $l_{_{2}}(z)$ in $R[z]/\langle z^{p^{s}}-1\rangle$ such that $$\big((z-1)^{a} + \gamma (z-1)^{t} g(z)\big)^{*} = l_{_{1}}(z) (z-1)^{a} + l_{_{1}}(z) \gamma (z-1)^{t} g(z) + l_{_{2}}(z) \gamma (z-1)^{b},$$ which implies 
     
       \begin{eqnarray}
        \label{eqn 1}
         (-1)^{a} (z-1)^{a} +  z^{a-t-k} \gamma (-1)^{t} (z-1)^{t} g^{*}(z) & = & l_{_{1}}(z) (z-1)^{a} + l_{_{1}}(z) \gamma (z-1)^{t} g(z) \nonumber \\
         & & + ~ l_{_{2}}(z) \gamma (z-1)^{b}.
        \end{eqnarray}
    
         Multiplying equation (\ref{eqn 1}) by $\gamma$, we get $$ \gamma (-1)^{a} (z-1)^{a} = \gamma l_{_{1}}(z) (z-1)^{a}.$$ On comparing degrees, we get that $l_{_{1}}(z)$ is a constant in $R[z]/\langle z^{p^{s}}-1\rangle$, i.e., $l_{_{1}}(z) = c + \gamma d$ for some $c,d \in F_{p^{m}}$. It follows that $$l_{_{1}}(z) = (-1)^{a} + \gamma d.$$ Substituting this value of $l_{_{1}}(z)$ in equation (\ref{eqn 1}), we get
     
       \begin{eqnarray}
         (-1)^{a} (z-1)^{a} +  z^{a-t-k} \gamma (-1)^{t} (z-1)^{t} g^{*}(z) & = & (-1)^{a} (z-1)^{a} + \gamma d (z-1)^{a} \nonumber \\
         & & +  (-1)^{a} \gamma (z-1)^{t} g(z) + l_{_{2}}(z) \gamma (z-1)^{b} \nonumber
       \end{eqnarray}
    
        which implies that
     
       \begin{eqnarray}
        (-1)^{a} (z-1)^{a} + z^{a-t-k} \gamma (z-1)^{t} B(z) & = & (-1)^{a} (z-1)^{a} + \gamma d (z-1)^{a} +  \gamma (z-1)^{t} A(z) \nonumber \\
         & & + ~ \gamma l_{_{2}}(z) (z-1)^{b}.
       \end{eqnarray}
    
        Since $(z-1)^{b} \lvert (z-1)^{a}$, we get that $$(z-1)^{b} \lvert \big( z^{a-t-k} \gamma (z-1)^{t} B(z) - \gamma (z-1)^{t}  A(z)\big).$$ Thus, $C$ is reversible implies
      
       \begin{equation}
         \label{eqn 3}
          (z-1)^{b-t} \lvert \big(z^{a-t-k} B(z) - A(z)\big).
       \end{equation}

        Conversely, let equation (\ref{eqn 3}) hold. Then there exists a polynomial $l_{_{3}}(z)$ in $R[z]$ such that
     
       \begin{equation}
        z^{a-t-k} B(z) - A(z)=l_{3}(z)(z-1)^{b-t} \nonumber
       \end{equation}
     
        which further implies that 
     
       \begin{equation}
        \label{eqn 4}
          z^{a-t-k} (-1)^{t} (z-1)^{t} g^{*}(z) = (-1)^{a} (z-1)^{t} g(z) + l_{_{3}}(z) (z-1)^{b}.
       \end{equation}
     
         Consider the generator $f(z)=(z-1)^{a} + \gamma (z-1)^{t} g(z).$ Then 
     
       \begin{equation*}
        \begin{split}
         f^{*}(z) & = (-1)^{a} (z-1)^{a} +  z^{a-t-k} \gamma (-1)^{t} (z-1)^{t} g^{*}(z)\\
         & =(-1)^{a} (z-1)^{a} + \gamma (-1)^{a} (z-1)^{t} g(z) + \gamma l_{_{3}}(z) (z-1)^{b} ~~~~~ \text{using equation (\ref{eqn 4}).}
        \end{split}
       \end{equation*}
       
        So, $f^{*}(z) \in C$. Also, $\big(\gamma(z-1)^{b}\big)^{*}=(-1)^{b}\gamma(z-1)^{b}$ belongs to $C.$ Using Lemma 2.4, we obtain that $C$ is reversible.
        
      \end{proof}

        Corollary 3.1 stated below is a particular case of Theorem 3.2.

      \begin{corollary}
   	    Let $C=\langle (z-1)^{a} + \gamma (z-1)^{t} g(z), \gamma (z-1)^{b}\rangle$ be a cyclic code having length $p^{s}$ over $R$. If $g(z)$ is a self reciprocal polynomial over $F_{p^{m}}$, then $C$ is reversible if and only if $(z-1)^{b-t} \lvert \big((-1)^{t} z^{a-t-k} - (-1)^{a}\big).$
   	  \end{corollary}

         Now, we give a few examples to illustrate our results. The Hamming distance and the cardinality of codes given in examples of this section have been calculated using the results given in \cite{bib4}.

       \begin{example}
    	 Let $C=\langle (z-1)^{3}+5(z-1)(3), 5(z-1)^{2}\rangle$ be a cyclic code having length $25$ over $Z_{_{25}}$ with  $d_{_{H}}(C)=2.$ 
    	 Also, $g(z)=3$, $a=3$, $b=2$, $t=1$ and $k=0$. Clearly $(z-1)^{b-t}=z-1$ and $\big((-1)^{t}z^{a-t-k}-(-1)^{a}\big)=-(z^{2}-1)$. Hence, $C$ satisfies the condition of Corollary 3.1. So, it is reversible.
       \end{example}
      
       \begin{example}
    	 Let $C=\langle (z-1)^{4}+\gamma z, \gamma (z-1)^{2}\rangle$ be a cyclic code having length 8 over $Z_{2}+\gamma Z_{2}$ having characteristic 2, $\gamma^{2}=0$ and $d_{_{H}}(C)=2.$ Here, $g(z)=z$, $a=4$, $b=2$, $t=0$ and $k=1$. Further, $(z-1)^{b-t}=(z-1)^{2}=z^{2}+1$ and $\big(z^{a-t-k}B(z)-A(z)\big)=z(z^{2}+1)$. Hence, $C$ satisfies the condition of Theorem 3.2. So, $C$ is reversible. 
       \end{example}

       \begin{example}
    	 Let $C=\langle (z-1)^{2}+2\gamma, \gamma(z-1) \rangle$ be a cyclic code having length 3 over $Z_{3}+\gamma Z_{3}$, $\gamma^{2}=0$ with $d_{_{H}}(C)=2.$ Here, $g(z)=2$ is a self reciprocal polynomial and $a=2$, $b=1$, $t=0$ and $k=0$. Clearly $(z-1)^{b-t}=z-1$ and $\big((-1)^{t}z^{a-t-k}-(-1)^{a}\big)=z-1$. Hence, $C$ satisfies the condition of Corollary 3.1 to be reversible. 
       \end{example}

       \begin{example}
  	    Consider the cyclic code $C=\langle (z-1)^{7}+3(z-1)\big(1+2(z-1)\big), 3(z-1)^4 \rangle$ of length $9$ over $Z_{_{9}}$ with $d_{_{H}}(C)=3.$ Here, $g(z)=2z+2$ is a self reciprocal polynomial and $a=7$, $b=4$, $t=1$ and $k=1$. Clearly $(z-1)^{b-t}=(z-1)^{3}$ and $\big((-1)^{t}z^{a-t-k}-(-1)^{a}\big)=2z^{5}-1$. Hence, $C$ does not satisfy the condition of Corollary 3.1. So, it is not reversible.
       \end{example}


       \begin{theorem}
     	 Let $C=\langle (z-1)^{a}+\gamma (z-1)^{t}g(z) \rangle$ be a cyclic code having length $p^{s}$ over $R$ such that $deg\big(g^{*}(z)\big) = deg\big(g(z)\big).$ Then $C$ is reversible implies that $char R=2.$
       \end{theorem}

      \begin{proof}
     	Let $C=\langle (z-1)^{a}+\gamma (z-1)^{t}g(z) \rangle$ be a cyclic code having length $p^{s}$ over $R$ such that $deg\big(g^{*}(z)\big) = deg\big(g(z)\big).$ Let $C$ be reversible, then by Theorem 3.2,
     	
        \begin{equation}
         \label{eqn 5}
          (z-1)^{a-t} \lvert \big( z^{a-t-k}(-1)^{t}g^{*}(z)-(-1)^{a}g(z) \big), \textit{where k = deg\big(g(z)\big).}
        \end{equation} 
         
         Further, $deg\big(g^{*}(z)\big) = deg\big(g(z)\big) = k$ implies that $deg\big( z^{a-t-k}(-1)^{t}g^{*}(z)-(-1)^{a}g(z) \big)= a-t.$ This together with Equation (\ref{eqn 5}) implies that  $z^{a-t-k}(-1)^{t}g^{*}(z)-(-1)^{a}g(z)$ is a constant multiple of $(z-1)^{a-t}.$\\
             
         Let us consider the following cases.\\
         \textbf{Case (a):} Both $a$ and $t$ are even integers.\\
         Then $a-t$ is also even and $(-1)^{a}=(-1)^{t}=1.$ The constant term and leading coefficient of $(z-1)^{a-t}$ are both equal to 1. The constant term and the leading coefficient of $z^{a-t-k}(-1)^{t}g^{*}(z)-(-1)^{a}g(z)$ are $-g(0)$ and $g(0)$ respectively. This is possible only if $char R=2.$\\\\
         \textbf{Case (b):} Both $a$ and $t$ are odd integers.\\
         In this case, the constant term and the leading coefficient of $(z-1)^{a-t}$ are both equal to 1. The constant term and the leading coefficient of $z^{a-t-k}(-1)^{t}g^{*}(z)-(-1)^{a}g(z)$ are $g(0)$ and $-g(0)$ respectively. This can hold only if $char R=2.$\\\\
         \textbf{Case (c):}  $a$ is even and $t$ is odd.\\
         In this case, the constant term and the leading coefficient of $(z-1)^{a-t}$ are $-1$ and $1$ respectively. The constant term and the leading coefficient of $z^{a-t-k}(-1)^{t}g^{*}(z)-(-1)^{a}g(z)$ are both equal to $-g(0)$. This can happen only if $char R=2.$\\\\
         \textbf{Case (d):}  $a$ is odd and $t$ is even.\\
         In this case, the constant term and the leading coefficient of $(z-1)^{a-t}$ are $-1$ and $1$ respectively. The constant term and the leading coefficient of $z^{a-t-k}(-1)^{t}g^{*}(z)-(-1)^{a}g(z)$ are both equal to $g(0)$. This is possible only if $char R=2.$\\
         
         It follows from all the above cases that if $C=\langle (z-1)^{a}+\gamma (z-1)^{t}g(z) \rangle$ is a cyclic code having length $p^{s}$ over $R$ such that $deg\big(g^{*}(z)\big) = deg\big(g(z)\big),$ then $C$ is reversible implies that $char R=2.$
         
       \end{proof}

         The following example shows that the above result does not hold when $deg\big( g^{*}(z)\big) < deg\big( g(z) \big).$

       \begin{example}
     	 Let $C=\langle (z-1)^{7}+3(z-1)g(z) \rangle,$ where $g(z)=z^{5}+z^{4}+z^{3}+z^{2}+z$ be a cyclic code having length 9 over the ring $Z_{9}$ having characteristic 9. It is easy to see that $deg\big( g^{*}(z)\big) < deg\big( g(z) \big).$ Also by using Theorem 3.2, we can see that $C$ is reversible. 
       \end{example}

      
        The lemma stated below due to A. Sharma and T. Sidana\cite{bib4} gives a characterization of all non-trivial MDS cyclic codes having length $p^{s}$ over a finite chain ring having nilpotency index 2.

       \begin{lemma}[\cite{bib4}]
     	Let $R=F_{_{p^{m}}}+\gamma F_{_{p^{m}}}$, $\gamma^{2}=0$ and $C$ be a cyclic code having length $p^{s}$ over $R$.\\
     	
     	\textbf{$(i)$} For $s=1$, all non-trivial distinct MDS cyclic codes having length $p$  over $R$ with respect to Hamming metric are $$\langle (z-1)^{a}+\gamma(z-1)^{t}g(z) \rangle,$$ where $1 \leq a \leq p-1,$ $g(z) \in F_{p^{m}}[z]$ is either zero or a unit in $R[z]/\langle z^{p^{s}}-1 \rangle$ with $deg\big(g(z)\big)<a-t$ and max$\{0,2a-p\} \leq t < a$ if $g(z) \neq 0.$\\ 
     	
     	\textbf{$(ii)$} For $s>1$, all non-trivial distinct MDS cyclic codes having length $p^{s}$ over $R$ with respect to Hamming metric are $$\langle z-1+\gamma g_{_{0}} \rangle ~ and ~ \langle (z-1)^{p^{s}-1}+\gamma (z-1)^{p^{s}-2}g_{_{0}} \rangle,$$ where $g_{_{0}} \in F_{_{p^{m}}}$.
       \end{lemma}
 
       \begin{remark}
       	 \label{remark1}
     	 It is obvious from Lemma 3.2 that a non-trivial MDS cyclic code having length $p^{s}$ over $R$ is principally generated by a monic polynomial.
       \end{remark}

       \begin{corollary}
       	Let $C$ 
       	be a non-trivial cyclic code having length $p^{s},$ $s>1$ over $R.$ If $C$ is MDS and reversible, then $char R=2.$
       \end{corollary}

       \begin{proof}
       	Let $C$ be a non-trivial MDS reversible cyclic code having length $p^{s},$ $s>1.$ Because $C$ is MDS, it follows from Lemma 3.2(ii) that $C=\langle z-1+\gamma g_{_{0}} \rangle$ or $C=\langle (z-1)^{p^{s}-1}+\gamma (z-1)^{p^{s}-2}g_{_{0}} \rangle$. Also $g(z)=g_{_{0}}$ is a constant and therefore, $deg\big(g^{*}(z)\big)=deg\big(g(z)\big).$ As $C$ is reversible and satisfies the condition of Theorem 3.3, it follows that $char R=2.$
       \end{proof}

       \begin{corollary}
      	Let $C=\langle (z-1)^{a}+\gamma (z-1)^{t}g(z), \gamma(z-1)^{b} \rangle$ be a non-trivial reversible cyclic code having length $p$ over $R$ such that $deg\big(g^{*}(z)\big) = deg\big(g(z)\big).$ Then, $C$ is MDS implies that $char R=2.$
       \end{corollary}
  
       \begin{proof}
       	It follows easily from Remark 3.1 and Theorem 3.3.
       \end{proof}



 
     
 
 

        Using the above results, we give a few examples of MDS reversible cyclic codes over some finite chain rings.
     
      \begin{example}
       Consider the cyclic code $C=\langle (z-1)^{255}+3\gamma (z-1)^{254} \rangle$ of length $256$ over the finite chain ring $R=F_{_{16}}+\gamma F_{_{16}}$, $\gamma^{2}=0$ with $char R=2$ and  $d_{_{H}}(C)=256.$ 
       Here, $g(z)=3$, $a=255$, $t=254$ and $k=0$. Further, $(z-1)^{a-t}=z-1$ and $\big(z^{a-t-k}(-1)^{t}-(-1)^{a}\big)=z+1=z-1$. Hence, $C$ satisfies the condition of Corollary 3.1. So, $C$ is reversible. Also, $\lvert C \lvert=2^{8}$ and $\lvert R \lvert=256.$ Since $\lvert C\lvert=\lvert R\lvert^{n-d_{H}(C)+1},$ $C$ is an MDS code over $R$.
      \end{example}

      \begin{example}
      	Consider the cyclic code $C=\langle z-1+\gamma \rangle$ of length $16$ over the finite chain ring $R=F_{_{4}}+\gamma F_{_{4}}$, $\gamma^{2}=0$ with $char R=2$ and  $d_{_{H}}(C)=2.$ 
      	Here, $g(z)=1$, $a=1$, $t=0$ and $k=0$. Clearly $(z-1)^{a-t}=z-1$ and $\big((-1)^{t}z^{a-t-k}-(-1)^{a}\big)=z+1=z-1$. Hence, $C$ satisfies the condition of Corollary 3.1. So, it is reversible. Also, $\lvert C \lvert=2^{60}$ and $\lvert R \lvert=16.$ Since $\lvert C\lvert=\lvert R\lvert^{n-d_{H}(C)+1},$ $C$ is an MDS code over $R$.
      \end{example}

\section{Reversible cyclic codes and their torsion codes}\label{sec4}

        Throughout this section, let $R$ be a finite commutative chain ring with $\langle \gamma\rangle$ its unique maximal ideal and $\nu$ the nilpotency index of $\gamma$. Let $F=R/\langle \gamma \rangle$ be the residue field of $R$. Let $C$ be a cyclic code having arbitrary length $n$ over $R$.\\

        In this section, the set of generators for a cyclic code $C$ of arbitrary length over $R$ as given in \cite{bib11} have been used to derive reversibility conditions on $C$. Further, some properties of torsion codes of $C$ have been explored. Also, the cardinality and Hamming distance of $C$ over $R$ have been determined.\\

        Let us now recall the structure of $C$ over $R$ as given in \cite{bib11} and state some useful results in this context.
   
        Let $f_{_{0}}(z),f_{_{1}}(z), \cdots, f_{_{m}}(z)$ be minimal degree polynomials in $C$ with leading coefficient of $f_{_{j}}(z)$ equal to $\gamma^{i_{_{j}}}u_{_{j}},$ where $u_{_{j}}$ is some unit in $R,$ $deg\big(f_{_{j}}(z) \big) < deg\big( f_{_{j+1}}(z) \big),$ $i_{_{j}}>i_{_{j+1}}$ and $i_{_{j}}$ is the smallest such power. If $i_{_{0}}=0,$ then $f_{_{0}}(z)$ is monic and we have $m=0.$
        

      \begin{lemma}[\cite{bib11}]
     	Let $C$ be a cyclic code having length $n$ over $R$ and $f_{_{j}}(z)$, $0 \leq j \leq m$ be polynomials as defined above. Then the following hold:\\
        $(i)$ $C$ is generated by the set $\{ f_{_{j}}(z) : j=0,1,\cdots,m \}$.\\
        $(ii)$ $f_{_{j}}(z) = \gamma^{i_{_{j}}} h_{_{j}}(z)$ for $0 \leq j \leq m$, where $h_{_{j}}(z)$ is a monic polynomial over the finite commutative chain ring with maximal ideal $\langle \gamma\rangle$ and nilpotency index $\nu - i_{_{j}}$.
       \end{lemma}

         The following results arise as an extension of results by J. Kaur et al. (\cite{bib8},\cite{bib7}) from cyclic codes having arbitrary length over Galois rings to cyclic codes having arbitrary length over finite chain rings.\\
         The theorem stated below gives necessary as well as sufficient conditions for $C$ to be a reversible cyclic code over $R$.

      \begin{theorem}
   	    Let $C$ be a cyclic code over $R$ generated by the polynomials ${f_{_{0}}(z),f_{_{1}}(z), \cdots, f_{_{m}}(z)}$ as defined earlier. Then $C$ is reversible if and only if\\
        $(i)$ $f^{*}_{_{0}}(z)=u_{_{0}}f_{_{0}}(z)$ for some unit $u_{_{0}} \in R$ and\\
        $(ii)$ $f^{*}_{_{r}}(z)-u_{_{r}}f_{_{r}}(z) \in \langle f_{_{s}}(z),f_{_{s-1}}(z), \cdots, f_{_{0}}(z)\rangle$ for some $s<r$ and a unit $u_{_{r}} \in R$, $0<r \leq \nu-1$.
      \end{theorem}

      \begin{proof}
        The proof follows on similar lines as that of Theorem 3.3.8 \cite{bib7}.
      \end{proof}

      \begin{lemma}
    	Consider a cyclic code $C$ of arbitrary length $n$ over a finite commutative chain ring $R$ generated by $\{f_{_{0}}(z), f_{_{1}}(z),\cdots,f_{_{m}}(z) \}$ as defined above, where $f_{_{j}}(z)=\gamma^{i_{_{j}}}h_{_{j}}(z)$ and $deg(h_{_{j}}(z))=t_{_{j}}$ for $j=0,1, \cdots, m$. Then for every $a(z)$ $\in$ Tor$_{_{i_{j}}}(C)$,  $deg\big(a(z)\big) \geq t_{_{j}}$.
      \end{lemma}

      \begin{proof}
    	The proof follows on the similar lines as that of Lemma 4.1 \cite{bib8}.
      \end{proof}

      \begin{theorem}
     	Let $C$ be a cyclic code over $R$ generated by  $\{f_{_{0}}(z), f_{_{1}}(z),\cdots,f_{_{m}}(z) \}$ as defined above. Then Tor$_{_{i_{j}}}(C)=\langle \overline{h_{_{j}}(z)}\rangle$ and $t_{_{j}}$ is the $i_{_{j}}$-th torsional degree of $C$.
      \end{theorem}
   
      \begin{proof}
       	It is easy to see that $\langle \overline{h_{_{j}}(z)}\rangle \subseteq$ Tor$_{_{i_{j}}}(C)$.\\
        Let $a(z)$ be an arbitrary element of Tor$_{_{i_{j}}}(C)$. By Lemma 4.2, $deg\big(a(z)\big) \geq t_{_{j}}$. Since $h_{_{j}}(z)$ is monic, $deg\big(\overline{h_{_{j}}(z)}\big)=deg\big(h_{_{j}}(z)\big)=t_{_{j}}$. Clearly $a(z)$ and $\overline{h_{_{j}}(z)}$ are polynomials over the residue field $F.$ So by division algorithm, there exist unique polynomials $q(z)$ and $r(z)$ in $F[z]/\langle z^{n}-1\rangle$ such that $$r(z)=a(z)-\overline{h_{_{j}}(z)}q(z)$$ where $r(z)=0$ or $deg\big(r(z)\big)<deg\big(\overline{h_{_{j}}(z)}\big)$. As $r(z) \in$ Tor$_{_{i_{j}}}(C)$, Lemma 4.2 implies that $r(z)=0$. Therefore, $a(z) \in \langle \overline{h_{_{j}}(z)}\rangle$. Hence,  Tor$_{_{i_{j}}}(C) = \langle \overline{h_{_{j}}(z)}\rangle$ and the $i_{_{j}}$-th torsional degree of $C$ is $deg\big(\overline{h_{_{j}}(z)}\big)=t_{_{j}}$.
      \end{proof}

      \begin{remark}
    	Let $C=\langle f_{_{0}}(z), f_{_{1}}(z),\cdots,f_{_{m}}(z)\rangle$ be a cyclic code having length $n$ over $R,$ where $f_{_{j}}(z)$ for $i=0,1,\cdots,m$ are polynomials as defined above. Clearly\\
        $(i)$~Tor$_{_{0}}(C)=$Tor$_{_{1}}(C)= \cdots =$Tor$_{_{i_{_{m}}-1}}(C)=\{0\}$\\
        $(ii)$~Tor$_{_{i_{_{j}}}}(C) =$ Tor$_{_{i_{_{j}}+1}}(C)= \cdots =$Tor$_{_{i_{_{j-1}}-1}}(C) \subset $Tor$_{_{i_{_{j-1}}}}(C) \ for \ j=1, 2,  \cdots, m$\\
        $(iii)$~Tor$_{_{i_{_{0}}}}(C) =$ Tor$_{_{i_{_{0}}+1}}(C)= \cdots =$ Tor$_{_{\nu-2}}(C)=$Tor$_{_{\nu-1}}(C).$
      \end{remark}

        The following lemma is required to determine the cardinality of $C$.

      \begin{lemma} [\cite{bib12}]
        If $C$ is a code over a finite commutative chain ring $R$, then $\lvert C \rvert=\prod_{j=0}^{\nu-1} \lvert $Tor$_{_{j}}(C) \rvert$.
      \end{lemma}

       \begin{theorem}
         Let $C$ be a cyclic code having arbitrary length $n$ over a finite chain ring having nilpotency index $\nu$ generated by polynomials $f_{_{0}}(z), f_{_{1}}(z),\cdots,f_{_{m}}(z)$ as defined earlier. If $\lvert F \rvert=p^{s}$, then
         $$\lvert C \rvert = p^{s\big(n\nu-(ni_{_{m}} + t_{_{0}}k_{_{0}}+t_{_{1}}k_{_{1}}+\cdots+t_{_{m}}k_{_{m}})\big)},$$
         where $t_{_{j}}$ for $j=0, 1, \cdots, m$ are the torsional degrees of Tor$_{_{i_{j}}}(C)$,  $k_{_{0}}=\nu-i_{_{0}}$ and $k_{_{j}}=i_{_{j-1}}-i_{_{j}}$ for $j=1,2,\cdots,m$.
       \end{theorem}

       \begin{proof}
   	     The result immediately follows from Remark 4.1, Lemma 4.3 and the fact that $\lvert $Tor$_{_{j}}(C) \rvert=p^{s(n-T_{_{j}})},$ where $T_{_{j}}$ is the degree of generator polynomial of Tor$_{_{j}}(C)$.\\
       \end{proof}

       \begin{theorem}
         Let $C$ be a reversible cyclic code over a finite commutative chain ring $R$. Then Tor$_{_{i}}(C)$ is a reversible cyclic code for each $i=0,1,\cdots,\nu-1$.
       \end{theorem}

       \begin{proof}
        The proof follows on the similar lines as that of Theorem 3.3.6 \cite{bib7}. However, we give the proof for the sake of completeness.\\
        Let $C$ be a cyclic code over $R.$ Then by Lemma 4.1(i), $C=\langle f_{_{0}}(z),f_{_{1}}(z), \cdots, f_{_{m}}(z) \rangle.$ By Remark 4.1(i), Tor$_{_{i}}(C)=\langle 0 \rangle$ for $0 \leq i \leq i_{_{m}}-1$ and hence Tor$_{_{i}}(C),$ for $0 \leq i \leq i_{_{m}}-1$ is reversible. By Theorem 4.2 and Remark 4.1(ii), there exists some $j$, $0 \leq j \leq m,$ such that Tor$_{_{i}}(C)=$ Tor$_{_{i_{_{j}}}}(C)=\langle \overline{h_{_{j}}(z)} \rangle$ for $i_{_{m}} \leq i \leq \nu-1.$ Therefore, $\gamma^{i_{_{j}}}\big( \overline{h_{_{j}}(z)}+\gamma A(z)\big) \in C$ for some $A(z) \in R[z]/\langle z^{n}-1 \rangle.$ As $C$ is reversible, $$ \gamma^{i_{_{j}}}\big(\overline{h_{_{j}}(z)}+\gamma A(z) \big)^{*} \in C.$$ It follows by using Lemma 2.3(ii) that $$\gamma^{i_{_{j}}}\big( \big(\overline{h_{_{j}}(z)}\big)^{*}+z^{t_{_{j}}-deg A(z)}\gamma A^{*}(z) \big) \in C.$$ So, $$\overline{\big( \big(\overline{h_{_{j}}(z)}\big)^{*}+z^{t_{_{j}}-deg A(z)}\gamma A^{*}(z) \big)} \in Tor_{_{i_{_{j}}}}(C)$$ which implies that $\big(\overline{h_{_{j}}(z)}\big)^{*} \in Tor_{_{i_{_{j}}}}(C).$ By Lemma 2.4, Tor$_{_{i_{_{j}}}}(C)$ is reversible.
        
       \end{proof}
    
         The example given below illustrates the fact that the converse of Theorem 4.4 is not true.

       \begin{example}
     	 Let $R = F_{_{3}} + \gamma F_{_{3}} + \gamma^{2} F_{_{3}},$ $\gamma^3=0.$ Let $C = \langle \gamma (z-1) + \gamma^{2}\rangle$ be a cyclic code of length $3$ over $R$. It can be observed that $\big(\gamma (z-1) + \gamma^{2}\big)^{*} \notin C$ thereby implying that $C$ is not reversible. However, all its torsion codes Tor$_{_{0}}(C) = \{ 0 \}$ and Tor$_{_{1}}(C) = $Tor$_{_{2}}(C) = \langle z-1 \rangle$ are reversible.
       \end{example}

       As the following proposition is an easy generalisation of [\cite{bib7}, Proposition 2.4.5], we omit the proof.
       
       \begin{proposition}
         Let $C=\langle f_{_{0}}(z), f_{_{1}}(z),\cdots,f_{_{m}}(z)\rangle$ be a cyclic code as defined above. Then $d_{_{H}}(C)=d_{_{H}}\big($Tor$_{_{i_{0}}}(C)\big)=d_{_{H}}\big(\langle\overline{h_{_{0}}(z)}\rangle\big).$
       \end{proposition}

        We present a few examples in support of the results given in this section.
   
       \begin{example}
        Consider the finite commutative chain ring $R = Z_{_{2}} + \gamma Z_{_{2}} + \gamma^{2} Z_{_{2}}$, $\gamma^{3} = 0$. All the 15 non-trivial cyclic codes of length 5 over $R$ have been listed by Abualrub in \cite{bib13}. Using Lemma 2.4, it is easy to see that all these non-trivial cyclic codes and their corresponding torsion codes are reversible.
       \end{example}

      \begin{example}
    	Consider the cyclic code $C=\langle (z-1)^{255}+3\gamma (z-1)^{254} \rangle$ of length $256$ over the finite chain ring $F_{_{16}}+\gamma F_{_{16}}$, where $\gamma^{2}=0$ and $char R=2.$ $C$ is reversible as shown in Example 3.6 above. Here, $Hamming$ distance of $C$ is 256. Its torsion codes Tor$_{_{0}}(C)=$Tor$_{_{1}}(C)=\langle (z-1)^{255} \rangle$ are MDS reversible cyclic codes over the residue field $F_{_{16}}$ with $Hamming$ distance equal to 256.
      \end{example}

      \begin{example}
    	Consider the cyclic code $C=\langle z-1+\gamma \rangle$ of length $16$ over the finite chain ring $R=F_{_{4}}+\gamma F_{_{4}}$, where $\gamma^{2}=0$ and $char R=2.$ $C$ is reversible as shown in Example 3.7 above. Here, $Hamming$ distance of $C$ is 2. Its torsion codes Tor$_{_{0}}(C)=$Tor$_{_{1}}(C)=\langle z-1 \rangle$ are MDS reversible cyclic codes over the residue field $F_{_{4}}$ with $Hamming$ distance equal to 2.
      \end{example}

\section{Conclusion}\label{sec5}

       In this paper, necessary and sufficient conditions for the reversibility of a cyclic code of arbitrary length over a finite commutative chain ring have been derived. MDS reversible cyclic codes of length $p^{s}$ over a finite chain ring of nilpotency index 2 have been characterized. A few examples of MDS reversible cyclic codes have been presented. Further, it is proved that the torsion codes of a reversible cyclic code over a finite chain ring are reversible. The cardinality and Hamming distance of a cyclic code over a finite commutative chain ring have also been determined.

\section*{Acknowledgments}

       The first author gratefully acknowledges the support provided by the Council of Scientific and Industrial Research (CSIR), India in the form of a research fellowship.

\section*{Declarations}

\begin{itemize}
	
	  \item Funding : This research is funded by the Council of Scientific and Industrial Research (CSIR), India in the form of research fellowship to the first author.
	  \item Financial or non-financial interests : The authors have no relevant financial or non-financial interests to disclose.
	  \item Conflict of interest : There is no Conflict of interest between the authors.
	  \item  Ethical responsibility : The manuscript in part or in full has not been submitted or published elsewhere.
	  \item Availability of data and materials : Data sharing is not applicable to this article as no datasets were generated or analysed during the current study.
	  \item Author's contributions : All of the authors declare that they have participated in the execution of the paper and approved the final version of the paper.

\end{itemize}

\end{document}